\documentstyle[twoside,fleqn,epsf,espcrc2]{article}

\newcommand{\emptyfigure}[2]{{\vbox to#2{\hbox to #1{\hfil} \vfil}}}

\title{
\vspace*{-26pt}
{\normalsize  \hfill UTHEP-291 \ \ \ } \\
\vspace*{-5pt}
{\normalsize  \hfill December 1994} \\
Finite Temperature Transition in Two Flavor QCD \\
with Renormalization Group Improved Action}

\author{
Y.~Iwasaki\address{Institute of Physics, University of Tsukuba,
                   Ibaraki 305, Japan},
K.~Kanaya$\mbox{}^{\rm a}$,
S.~Sakai\address{Faculty of Education, Yamagata University,
                 Yamagata 990, Japan}
and
T.~Yoshi\'e$\mbox{}^{\rm a}$\thanks{Talk presented by T.Yoshi\'e at 
{\it Lattice 94}.}
}
       
\begin{document}

\begin{abstract}
The finite temperature transition or crossover in QCD 
with two degenerate Wilson quarks
is investigated using a renormalization group improved action.
At $\beta=2.0$ and 2.1 where $a^{-1} \sim 1.0-1.2$ GeV,
the expectation value of the Polyakov loop and the pion screening mass
on an $8^3 \times 4$ lattice vary smoothly with the hopping parameter 
through the transition/crossover.
The quark screening mass in the high temperature phase 
agrees well with that in the low temperature phase
calculated on an $8^4$ lattice.
The smooth transition of the observables is totally different 
from the sharp transition found for the standard action at $\beta=5.0$ and 5.1
where $a^{-1}$ is also $1.0-1.2$ GeV.
\end{abstract}

\maketitle

\section{Introduction}
Recently MILC group has extensively investigated the finite
temperature transition/crossover 
in the case of two degenerate Wilson fermions
and found unexpected phenomena\cite{MILC4,MILC6}.
For $N_t$=4, when the quark mass is heavy the transition is smooth and
when the quark mass is light it is also smooth. However, in the
intermediate mass region($\beta \sim 5.0$), 
the transition is very sharp. This is completely opposite
to what was supposed to be realized:
It was supposed to become weaker or disappear.
Furthermore, for $N_t=6$, in  the range of
intermediate mass, they observed clear two
state signals.

\begin{figure*}[t]
\begin{center}
\leavevmode
\epsfysize=290pt
  \epsfbox{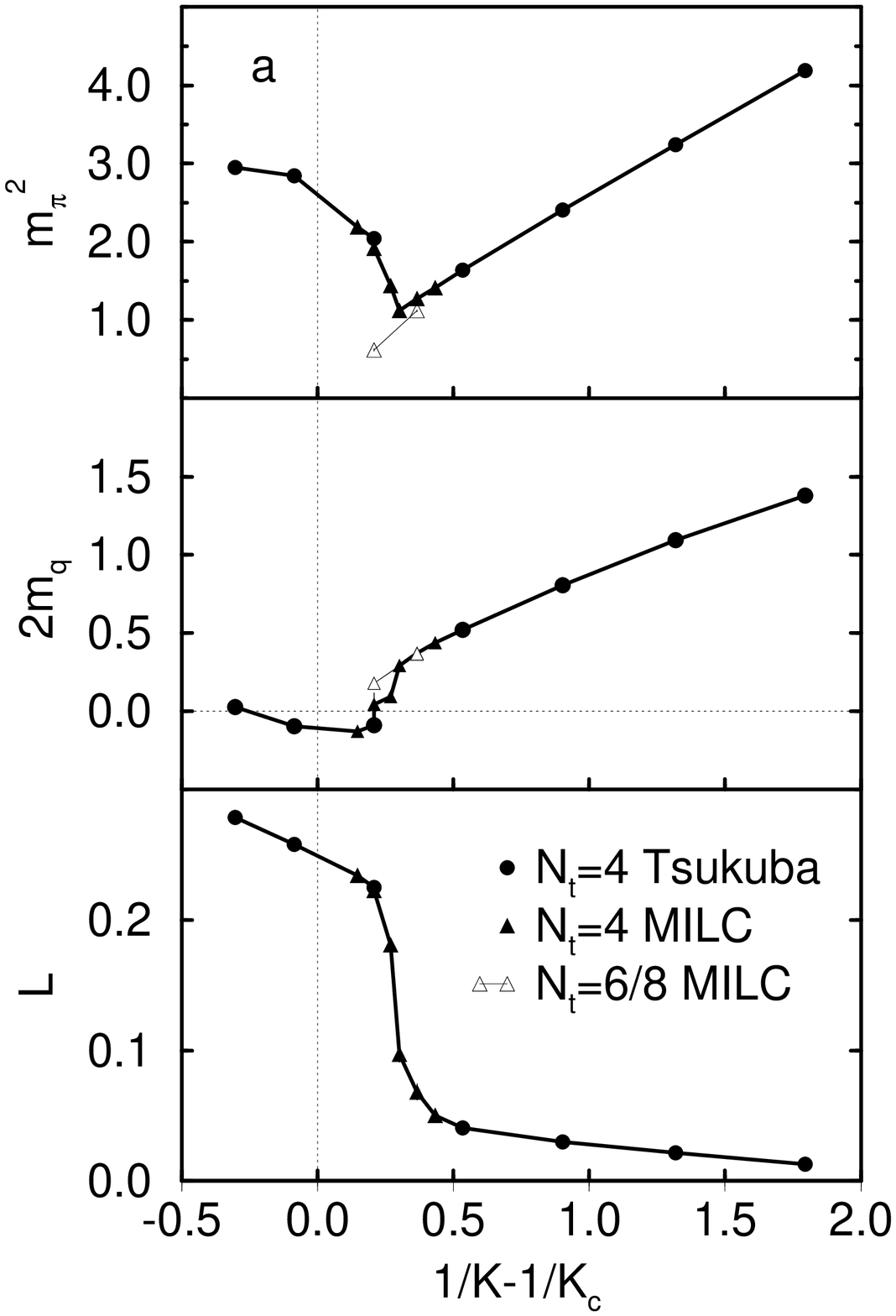}
\leavevmode
\epsfysize=290pt
  \epsfbox{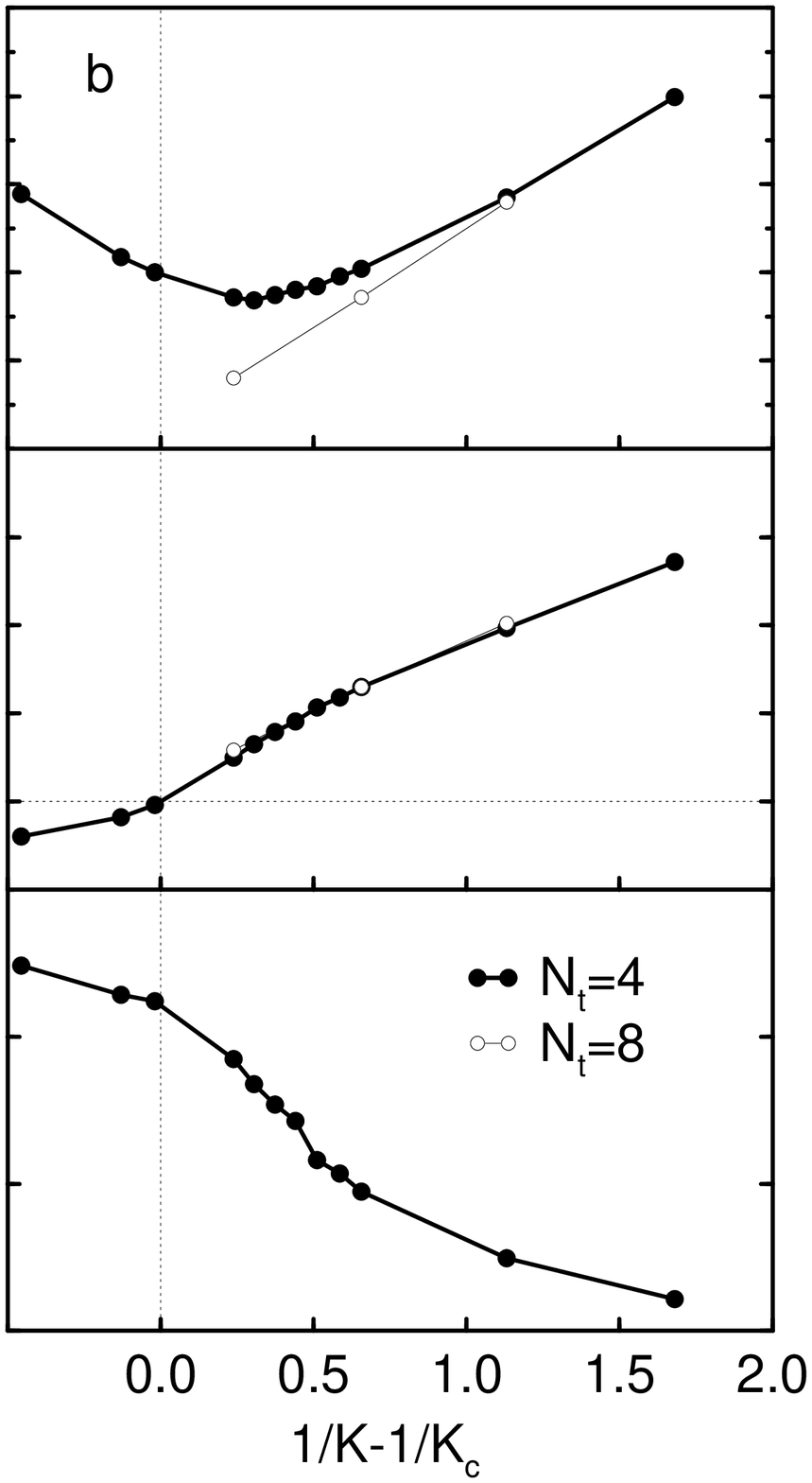}
\end{center}
\vskip -14mm
\caption{
Observables on $N_t=4$ (filled symbols) and $N_t=6$ or 8 (open symbols) lattice.
a) At $\beta=5.0$ with the standard action. Data are taken from
\protect\cite{MILC4} and \protect\cite{T0}.
b) At $\beta=2.0$ with the improved action.} 
\label{fig}
\end{figure*}

We reproduce in fig.\ref{fig}-a the data\cite{MILC4}
for the pion screening mass squared $m_{\pi}^2$,
the expectation value of the Polyakov loop $L$ and
the quark screening mass $m_q$ at $\beta=5.0$, 
together with our data\cite{T0}.
The transition of the Polyakov loop is sharp and 
a cusp appears in the pion screening mass at the transition point.
The quark mass defined through an axial-vector Ward identity\cite{ItohNP,Bo}
also exhibits peculiar behavior.
The $m_q$ in the high temperature phase on the $N_t=4$ lattice
behaves singularly and
does not agree with that in the low temperature phase.
This is in clear contrast with the fact that
the $m_q$ is almost independent of whether the system is in 
the high temperature phase or in the low temperature phase
at $\beta=5.85$ in the quenched QCD \cite{ChQ} and
at $\beta=5.5$ for $N_f=2$ case\cite{ChNf2}. 

One possible reason for these phenomena is that we are far from the continuum
limit. 
In this work, we study 
the finite temperature transition/crossover
in the case of two degenerate quarks
using a renormalization group (RG) improved action where
we are supposed to be close to the continuum limit.
Our main concern is to see whether the unexpected phenomena
described above
persist with the improved action.

\section{Improved action}
The RG improved action for pure gauge theory we proposed about ten years
ago is
\[
S_g = {1/g^2} \{c_0 \sum ({\rm plaquette}) 
               + c_1 \sum (1\times 2 {\rm \ loop})\}
\]
with $c_1=-0.331$ and $c_0=1-8c_1$. 
Each oriented loop appears once in the sum.
The form of the gauge action is determined
by a block-spin RG analysis in the pure gauge theory\cite{IM}.
We calculated the string tension\cite{IMst,IMscaling} 
and the hadron spectrum\cite{ItohNP,IMhd}
and investigated the topological properties and the $U(1)$ problem\cite{IMU1}
in the quenched QCD with this improved action.
We observe the following among others:
The approach to the perturbative result
for the expectation value of the plaquette
is much faster than the case of the
standard one-plaquette action\cite{IMscaling}.
Also note that the scale parameter of the 
improved action $\Lambda_{IM}$
is close to the $\Lambda_{\overline{MS}}$:
$\Lambda_{\overline{MS}}/\Lambda_{IM} \sim$ 0.488\cite{IML}.
Therefore the bare coupling of the improved action is already an
improved coupling\cite{LM}.

With these results it may be worthwhile to use a RG improved
action for full QCD.
The action we use as an improved action for full QCD is given by
$$S= S^{IM}_{gauge}+S^{Wilson}_{quark}.$$
The action is a sum of the RG improved action for gluons
and the Wilson action for quarks.
This action reduces to the improved pure gauge action
in the heavy quark mass limit. 
There is also a possibility to improve the quark action.
However we take the Wilson action for the quark action as a first step,
because we believe that the effect of the improvement of the gauge sector
is much more significant than that of the quark sector by the following
reasoning:
After integration over fermion variables,
the quark action may be written as a sum over various types of Wilson loops,
assuming the convergence of hopping parameter expansion.
The coefficients of various types of Wilson loops thus obtained are
much smaller compared with $c_1=-0.331$ for the improved pure gauge action.

\section{Numerical Simulations} 
Numerical simulations are done 
at $\beta=$ 2.0 and 2.1 
on an $8^3 \times 4$ lattice 
at several hopping parameters.
In order to fix the scale and to compare the results 
in the high temperature phase with those in the low temperature phase,
we also make simulations on an $8^3 \times 8$ lattice.
We use the hybrid Monte Carlo algorithm to generate gauge configurations
with molecular dynamics step size 
$\Delta \tau$ = 0.01. Momentum refresh is done at every one unit of
simulation time.
Acceptance rates are larger than 0.92 for all cases.
We discard first 100 $\sim$ 400 trajectories for thermalization and 
use 100 $\sim$ 1000 trajectories for measurement.
The numbers of trajectories for thermalization and measurement are chosen
taking into account 
whether the simulation point is close or not to the transition point.
Wilson loops and Polyakov loops are measured at every trajectory.
Screening correlation functions are measured at every 5 trajectories on an
$8^2 \times 16 \times N_t$ lattice obtained by doubling the original lattice.

Using the $\rho$ meson mass calculated on the $8^3 \times 16$ lattice
as an input,
we obtain $a^{-1} = $ 1.01 GeV and 1.18 GeV at $\beta=$ 2.0 and 2.1, 
respectively.
Therefore these $\beta$'s correspond approximately to $\beta =$ 5.0 
and 5.1 of the standard action, respectively.

\section{Results} 

Fig.\ref{fig}-b shows the results for the observables at
$\beta =$ 2.0 versus $1/K-1/K_c$,
where the $K_c$ is determined from the quark mass on the $N_t=4$ lattice.
The data indicate the following:
1) The transition/crossover of the Polyakov loop 
around $1/K-1/K_c \sim 0.5$ is very smooth.
2) The hopping parameter dependence of the quark mass is also smooth and
its value on the $N_t =4$ lattice agrees well with
that on the $N_t = 8$ lattice. This implies that 
the value of the quark mass is almost independent
of whether the system is in the low temperature phase 
or in the high temperature phase.
3) The pion screening mass 
does not exhibit a cusp.

These features of the transition/crossover at $\beta=2.0$ are 
totally different from those observed
for the standard action at $\beta = 5.0$. 
Note that $a^{-1}$ are about 1 GeV for both cases.
We also observe a similar very clear contrast between the results 
at $\beta=2.1$ with the improved action and 
those at $\beta=5.1$ with the standard action
where $a^{-1}$ are both approximately 1.2 GeV.

Furthermore
preliminary results in the range of of $\beta=1.6-2.15$
indicate that
the transitions are smooth from $\beta=2.15$ down to at least $\beta=1.8$.
Therefore we conclude 
that the sharp transition in the intermediate mass
region disappears with the improved action at $N_t=4$.
This suggests a strong possibility that the first order phase transition
observed by MILC at $N_t=6$ disappears with this action.
To confirm this, we are planning to extend the work to a wider range
of parameter, in particular, on a lattice with $N_t=6$.

The simulations are performed with Fujitsu VPP500/30 at
University of Tsukuba.
We would like to thank C.~DeTar for providing us with
data by MILC collaboration.
This work is in part supported by the Grant-in-Aid
of Ministry of Education, Science and Culture
(No.06NP0601).

\end{document}